\documentclass[12pt]{article}
\usepackage{amsmath}
\usepackage{times}
\usepackage{graphicx}
 \usepackage{setspace}

\usepackage[authoryear]{natbib}
\usepackage{rotating}
\usepackage{bbm}
\usepackage{latexsym}

\usepackage{amssymb}
\usepackage{caption}
\usepackage{subcaption}
\usepackage{url}
\usepackage{booktabs}
\usepackage[dvipsnames,table]{xcolor}
\usepackage{longtable}

\usepackage{todonotes}
\usepackage{adjustbox}
\usepackage{tabularx}
\usepackage{color}

\definecolor{light-gray}{gray}{0.9}

\usetikzlibrary{matrix}
\newcommand{\captionfonts}{\normalsize}

\makeatletter  
\long\def\@makecaption#1#2{%
  \vskip\abovecaptionskip
  \sbox\@tempboxa{{\captionfonts #1: #2}}%
  \ifdim \wd\@tempboxa >\hsize
    {\captionfonts #1: #2\par}
  \else
    \hbox to\hsize{\hfil\box\@tempboxa\hfil}%
  \fi
  \vskip\belowcaptionskip}
\makeatother   
\begin{document}
\hspace{13.9cm}1

\ \vspace{20mm}\\

{\LARGE Replay in Deep Learning: Current Approaches and Missing Biological Elements}

\ \\
{\bf \large Tyler L. Hayes$^{\displaystyle 1}$, Giri P. Krishnan$^{\displaystyle 2}$, Maxim Bazhenov$^{\displaystyle 2}$, Hava T. Siegelmann$^{\displaystyle 3}$, Terrence J. Sejnowski$^{\displaystyle 2,4}$, Christopher Kanan$^{\displaystyle 1,5,6}$}\\
{$^{\displaystyle 1}$Rochester Institute of Technology, Rochester, NY, USA}\\
{$^{\displaystyle 2}$University of California at San Diego, La Jolla, CA, USA}\\
{$^{\displaystyle 3}$University of Massachusetts, Amherst, MA, USA}\\
{$^{\displaystyle 4}$Salk Institute for Biological Studies, La Jolla, CA, USA}\\
{$^{\displaystyle 5}$Paige, New York, NY, USA}\\
{$^{\displaystyle 6}$Cornell Tech, New York, NY, USA}\\


\thispagestyle{empty}
\markboth{}{NC instructions}
\ \vspace{-0mm}\\
%
%Abstract
\begin{center} {\bf Abstract} \end{center}
    Replay is the reactivation of one or more neural patterns, which are similar to the activation patterns experienced during past waking experiences. Replay was first observed in biological neural networks during sleep, and it is now thought to play a critical role in memory formation, retrieval, and consolidation. Replay-like mechanisms have been incorporated into deep artificial neural networks that learn over time to avoid catastrophic forgetting of previous knowledge. Replay algorithms have been successfully used in a wide range of deep learning methods within supervised, unsupervised, and reinforcement learning paradigms. In this paper, we provide the first comprehensive comparison between replay in the mammalian brain and replay in artificial neural networks. We identify multiple aspects of biological replay that are missing in deep learning systems and hypothesize how they could be utilized to improve artificial neural networks.
%%%%%%%%%%%

%% main text
%%%%%%%%%%%%%%%%%%%%%%%%%%%%%%%%%%%%%%%%%%%%%%%%%%%%%%%%%%%%%%%%%%%%%%%%%%%%%%%% 
\section{Introduction}
\label{sec:intro}

While artificial neural networks now rival human performance for many tasks, the dominant paradigm for training these networks is to train them once and then to re-train them from scratch if new data is acquired. This is wasteful of computational resources, and many tasks involve updating a network over time. However, standard training algorithms, i.e., online error backpropagation, produce catastrophic forgetting of past information when trained from a non-stationary input stream, e.g., incrementally learning new classes over time or most reinforcement learning problems~\citep{Abraham2005Memory,robins1995catastrophic}. The root cause of catastrophic forgetting is that learning requires the neural network's weights to change, but changing weights critical to past learning results in forgetting. This is known as the stability-plasticity dilemma, which is an important problem in deep learning and neuroscience~\citep{Abraham2005Memory}.  

In contrast, humans can continually learn and adapt to new experiences throughout their lifetimes and rarely does learning new information cause humans to catastrophically forget previous knowledge~\citep{french1999catastrophic}. In the mammalian brain, one mechanism used to combat forgetting and facilitate consolidation is replay - the reactivation of past neural activation patterns\footnote{For simplicity, we use replay to refer to both reactivation and replay. In the neuroscience literature, replay typically refers to the reactivation of a sequence of \emph{more than one} neural patterns in the same sequence they occurred during waking experience, but here we define it as the reactivation of \emph{one or more} neural patterns.}~\citep{mcclelland1995there,kumaran2016learning,mcclelland2020integration}. Replay has primarily been observed in the hippocampus, which is a brain structure critical for consolidating short-term memory to long-term memory. Replay was first noted to occur during slow-wave sleep, but it also occurs during Rapid Eye Movement (REM) sleep~\citep{louie_temporally_2001,Eckert2020Neural,Kudrimoti1999Reactivation} and while awake, potentially to facilitate the retrieval of recent memories~\citep{walker2004sleep}.

%% ARTIFICIAL VS BIOLOGICAL NETWORKS FIGURE %%
\begin{figure}[htb!]
     \centering
     \begin{subfigure}[b]{0.3\textwidth}
         \centering
         \includegraphics[width=\textwidth]{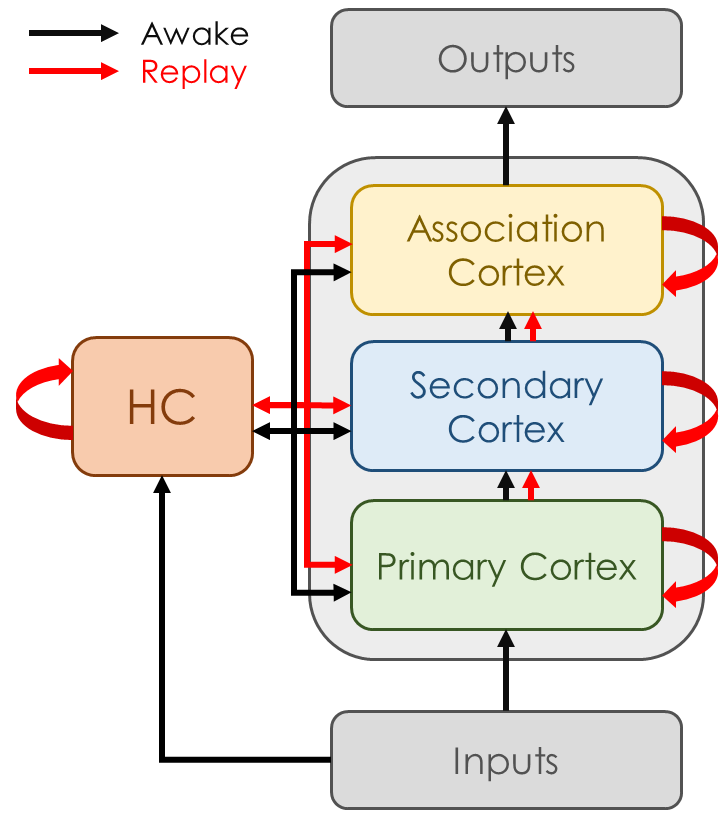}
         \caption{Replay in Biological Neural Networks}
         \label{fig:biological-replay}
     \end{subfigure}
     \hfill
     \begin{subfigure}[b]{0.3\textwidth}
         \centering
         \includegraphics[width=0.94\textwidth]{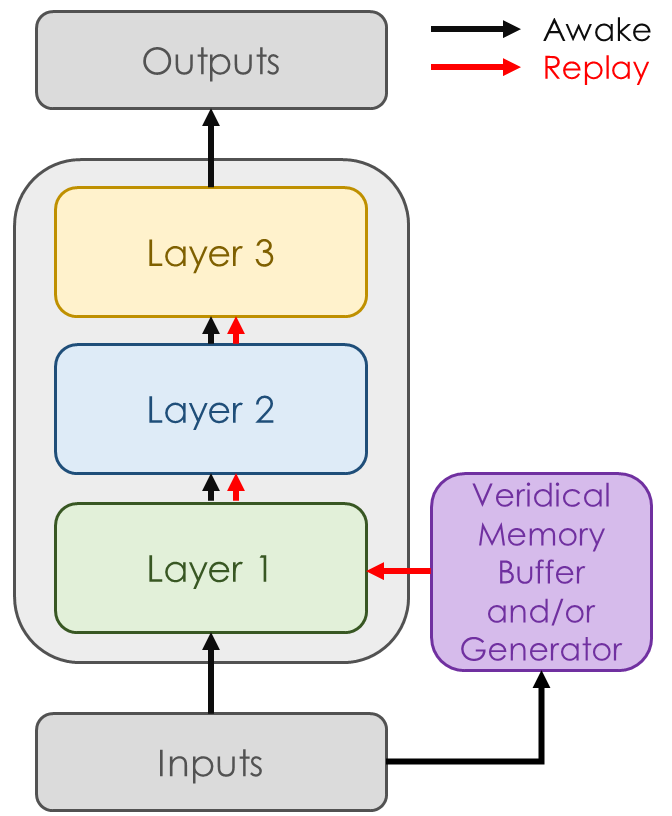}
         \caption{Veridical Replay in ANNs}
         \label{fig:artificial-replay-veridical}
     \end{subfigure}
     \hfill
     \begin{subfigure}[b]{0.3\textwidth}
         \centering
         \includegraphics[width=0.94\textwidth]{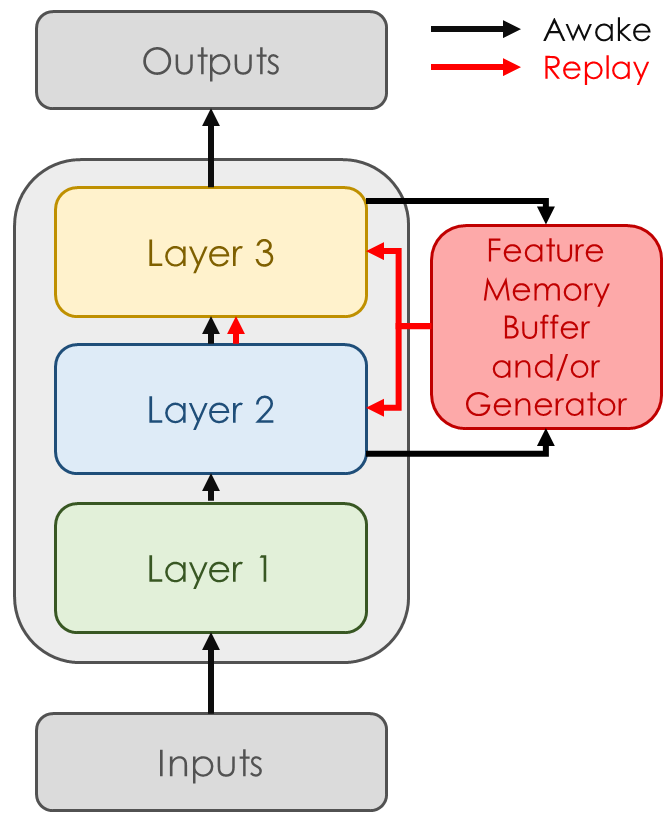}
         \caption{Representational Replay in ANNs}
         \label{fig:artificial-replay-generative}
     \end{subfigure}
        \caption{High-level overview of the flow of learning activity during awake and replay stages in biological networks versus artificial neural networks (ANNs). While replay occurs in several brain regions both independently and concurrently, replay in most artificial implementations occurs concurrently from a single layer. While the hippocampal complex (HC) can be used for both replay and inference in biological networks, the memory buffers in artificial replay implementations are mostly used to train the neural network that makes predictions. Figure (a) illustrates replay in biological networks. Figures (b) and (c) are examples of replay in an ANN with 3 hidden layers. For networks with more layers, the layer for representational replay can be chosen in a variety of ways (see text). 
        }
        \label{fig:biological-versus-ann-replay}
\end{figure}

In artificial networks, the catastrophic forgetting problem during continual learning has been successfully addressed by methods inspired by replay~\citep{rebuffi2016icarl,castro2018end,wu2019large,hou2019unified,hayes2019remind}. In the most common implementation, replay involves storing a subset of previous veridical inputs (e.g., RGB images) and mixing them with more recent inputs to update the networks~\citep{rebuffi2016icarl,castro2018end,wu2019large,hou2019unified,andrychowicz2017hindsight,Schaul2016PrioritizedER,lesort2019generative,wu2018memory,draelos2017neurogenesis}. This preserves representations for processing previous inputs while enabling new information to be learned. In contrast, the brain replays highly processed representations of past inputs, e.g., those stored within the hippocampus~\citep{Teyler2007The}, and a similar approach has been used to enable continual learning for artificial networks in some recent works that replay high-level feature representations~\citep{hayes2019remind,iscen2020memory,caccia2019online,pellegrini2019latent,van2020brain}. While this is closer to biology, many facets of replay in biology have not been incorporated into artificial networks, but they could potentially improve generalization, abstraction, and data processing. A high-level depiction of the differences between biological and artificial replay is shown in Fig.~\ref{fig:biological-versus-ann-replay}.

%%%% MAIN TABLE CONCLUSIONS %%%%
\begin{table}%[t]
    \caption{High-level overview of replay mechanisms in the brain, their hypothesized functional role, and their implementation/use in deep learning.
    }
    \label{tab:main-table}
    \centering
\rowcolors{0}{white}{light-gray}
{\small %
\begin{tabular}{|p{.3\textwidth}|p{.3\textwidth}|p{.3\textwidth}|}
\toprule\rowcolor{white}
\textbf{\textsc{Replay Mechanism}} & \textbf{\textsc{Role in Brain}}  & \textbf{\textsc{Use in Deep Learning}} \\
\midrule
Replay includes contents from both new and old memories & Prevents forgetting & Interleave new data with old data to overcome forgetting \\
Only a few selected experiences are replayed & Increased efficiency, weighting experiences based on internal representation & Related to subset selection for what should be replayed \\
Replay can be partial (not entire experience) & Improves efficiency, allows for better integration of parts, generalization, and abstraction & Not explored in deep learning \\
Replay observed at sensory and association cortex (independent and coordinated) & Allows for vertical and horizontal integration in hierarchical memory structures & Some methods use representational replay of higher level inputs or feature maps \\
Replay modulated by reward & Allow reward to influence replay & Similar to reward functions in reinforcement learning \\
Replay is spontaneously generated (without external inputs) & Allows for all of the above features of replay without explicitly stored memories & Some methods replay samples from random inputs \\
Replay during NREM is different than replay during REM & Different states allow for different types of manipulation of memories & Deep learning currently focuses on NREM replay and ignores REM replay \\
Replay is temporally structured & Allows for more memory combinations and follows temporal waking experiences & Largely ignored by existing methods that replay static, uncorrelated inputs \\
Replay can happen in reverse & Allow reward mediated weighting of replay & Must have temporal correlations for reverse replay \\
Replay is different for novel versus non-novel inputs & Allows for selective replay to be weighted by novelty & Replay is largely the same independent of input novelty \\
\bottomrule
\end{tabular}
}
\end{table}

In this paper, we first describe replay's theorized role in memory consolidation and retrieval in the brain, and provide supporting evidence from neuroscience and psychology studies. We also describe findings in biology that deviate from today's theory. Subsequently, we discuss how replay is implemented to facilitate continual learning in artificial neural networks. While there have been multiple reviews of replay in the brain~\citep{Tingley2020On,olafsdottir2018role,pfeiffer2020content,foster2017replay,robertson2020memories} and reviews of continual learning in artificial networks~\citep{kemker2018forgetting,parisi2019continual,de2019continual,belouadah2020comprehensive}, we provide the first comprehensive review that integrates and identifies the gaps between replay in these two fields. While it is beyond the scope of this paper to review everything known about the biology of replay, we highlight the salient differences between known biology and today's machine learning systems to help biologists test hypotheses and help machine learning researchers improve algorithms. An overview of various replay mechanisms in the brain, their hypothesized functional role, and their implementation in deep neural networks is provided in Table~\ref{tab:main-table}.

%%%%%%%%%%%%%%%%%%%%%%%%%%%%%%%%%%%%%%%%%%%%%%%%%%%%%%%%%%%%%%%%%%%%%%%%%%%%%%%%%%%%%%%%%%%%%%%%%%%%%%%%%%%%%%%%
\section{Replay in Biological Networks}
\label{sec:replay-biology}

Memory in the brain is the process of encoding, storing, and retrieving information. Encoding involves converting information into a format that can be stored in short-term memory, and then a subset of short-term memories are consolidated for long-term storage. Consolidation is a slow process that involves the integration of new memories with old~\citep{mcgaugh2000memory}. Splitting learning into short-term and long-term memory allows the brain to efficiently solve the stability-plasticity problem. The consolidation phase is used for long-term storage of declarative, semantic, and procedural memories~\citep{rasch2013sleep, born2010slow, stickgold2005sleep, stickgold_sleep:_2012}.

Consolidation occurs during periods of rest or sleep, where spiking activity during replay initiates long-term changes in synapses through activity dependent plasticity processes. Consolidation is well understood for declarative and semantic memory, which depend on the hippocampus. Bilateral removal of the hippocampus results in anterograde amnesia and the inability to form new semantic memories~\citep{nadel1997memory}. The primary input to the hippocampus is the entorhinal cortex, which receives highly processed information from all sensory modalities and the prefrontal cortex. While the hippocampus allows for the quick assimilation of new information, medial prefrontal cortex is used for long-term storage of memories and generalization~\citep{bontempi1999time}. These generalization capabilities are a result of medial prefrontal cortex using a slower learning rate and densely encoding memories with overlapping representations, whereas the hippocampus uses a faster learning rate in conjunction with sparsity and indexing mechanisms~\citep{Teyler2007The}. Hebbian learning and error-driven schemes are used by both the hippocampus and medial prefrontal cortex~\citep{o2001conjunctive}. Initially, the hippocampus is used to retrieve new memories, but over time medial prefrontal cortex is instead used for retrieval~\citep{kitamura2017engrams}. Results from \citep{kitamura2017engrams} suggest that when new memories are created, neurons are allocated in both the hippocampus and medial prefrontal cortex, with unused neurons in the hippocampus engaged immediately for formation and neurons in medial prefrontal cortex being epigenetically `tagged' for later storage~\citep{lesburgueres2011early,bero2014early}. This switch from hippocampus to cortex occurs over a period of several days with several episodes of sleep, however, it can take months or even years to make memories completely independent of the hippocampus. Replay during sleep determines which memories are formed for long-term storage. Moreover, while replay contributes to long-term memory consolidation, the associated retrieval processes have been shown to change memories qualitatively~\citep{jonker2018neural}. These qualitative changes are thought to strengthen the representations of episodically similar memories and have been underexplored in computational models. 

The complementary learning systems (CLS) theory describes long-term memory consolidation based on the interplay between the hippocampus and the neocortex~\citep{mcclelland1995there,kumaran2016learning}. In this framework, the hippocampus and cortex act complimentary to one another. The hippocampus quickly learns short-term instance-level information, while the cortex learns much more slowly and is capable of better generalization. The CLS theory~\citep{mcclelland1995there,kumaran2016learning} also suggests that the brain generalizes across a variety of experiences by retaining episodic memories in the hippocampal complex and consolidating this knowledge to the neocortex during sleep.

Consolidation of hippocampus independent short-term memory, such as emotional and procedural memory, is also enhanced by sleep~\citep{mcgaugh2000memory,Hu2006Sleep,Payne2008Sleep,Wagner2001Emotional}. Sleep improves motor sequence learning~\citep{walker2005sleep, walker2002practice}, motor adaptation~\citep{stickgold2005sleep}, and goal-related sequence tasks~\citep{Albouy2013Daytime,Cohen2005Off-line}. Learning motor tasks involves many brain regions including motor cortex, basal ganglia, and hippocampus~\citep{Debas2010Brain}. While some improvement in sequential motor tasks may arise from the hippocampal contribution during sleep~\citep{King2017Sleepinga}, improvement in motor adaptation tasks does not involve the hippocampus~\citep{Debas2010Brain}. Sleep has also been shown to prevent interference between procedural and declarative tasks~\citep{Brown2007Off-Line}, suggesting a role for sleep in preventing interference during consolidation of different memory types. 

\begin{figure}%[th]
     \centering
     \begin{subfigure}[b]{\textwidth}
         \centering
         \includegraphics[width=\textwidth]{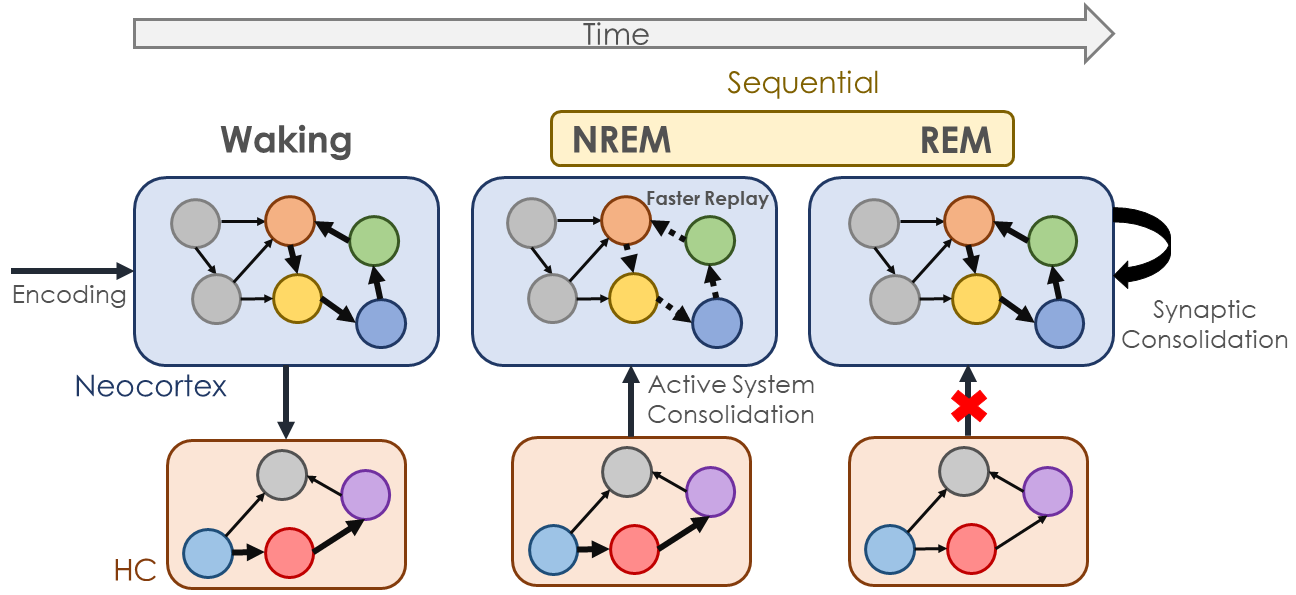}
         \caption{Depiction of the contributions of replay to memory formation and consolidation during waking, NREM, and REM stages.}
         \label{subfig:replay-stages}
     \end{subfigure}\\
     \begin{subfigure}[b]{0.24\textwidth}
         \centering
         \includegraphics[width=\textwidth]{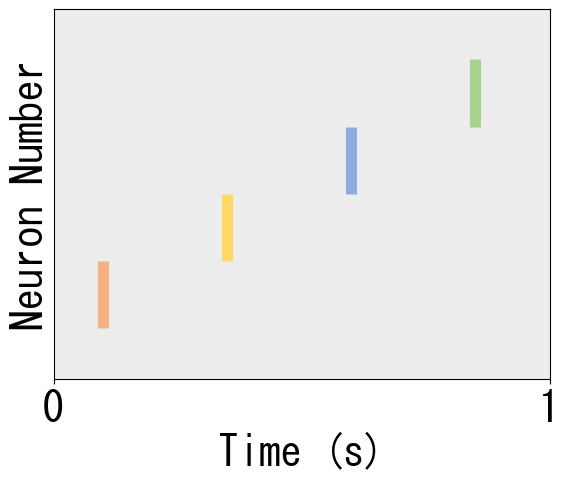}
         \caption{Awake Activity}
         \label{subfig:awake-spike}
     \end{subfigure}
     \hfill
     \begin{subfigure}[b]{0.24\textwidth}
         \centering
         \includegraphics[width=\textwidth]{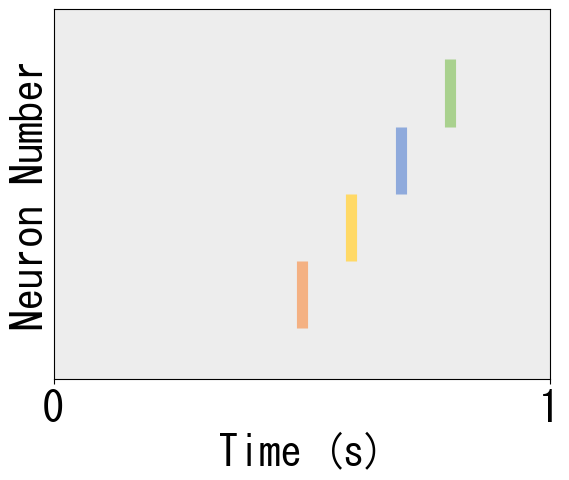}
         \caption{Faster Replay}
         \label{subfig:fast-spike}
     \end{subfigure}
     \hfill
     \begin{subfigure}[b]{0.24\textwidth}
         \centering
         \includegraphics[width=\textwidth]{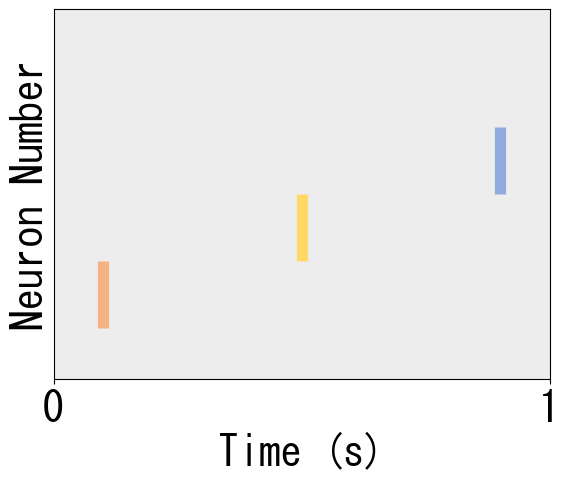}
         \caption{Partial Replay}
         \label{subfig:partial-spike}
     \end{subfigure}
     \hfill
     \begin{subfigure}[b]{0.24\textwidth}
         \centering
         \includegraphics[width=\textwidth]{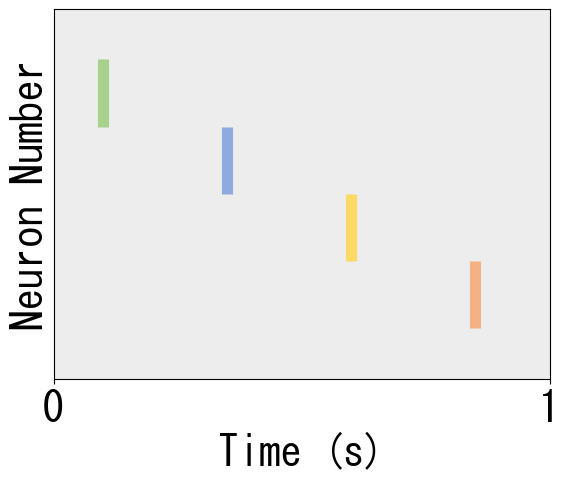}
         \caption{Reverse Replay}
         \label{subfig:reverse-spike}
     \end{subfigure}
        \caption{(a) Visualization of the contribution of replay in the hippocampal complex (HC) and the neocortex during different stages. Specifically, during waking hours, experiences are encoded in the HC and neocortex. While asleep, humans cycle between NREM and REM sleep stages, with NREM stages getting shorter and REM stages getting longer as the night progresses. In NREM, recent experiences are consolidated. In REM, internal experiences are organized. We also illustrate spike traces of neocortical outputs during (b) awake activity, (c) faster replay (NREM), (d) partial replay (NREM and REM), and (e) reverse replay (reinforcement learning). Note that activity during REM has been observed to be similar to that during waking experiences.}
        \label{fig:biological-replay-types}
\end{figure}

Different sleep stages have distinct roles in memory consolidation. Non-rapid eye movement (NREM) sleep is strongly associated with consolidation of declarative memory~\citep{Diekelmann2014Sleep, walker2010overnight}. In contrast, rapid eye movement (REM) sleep promotes the organization of internal representations~\citep{dumay2007sleep,haskins2008perirhinal,bader2010recognition,tibon2014associative}, abstraction~\citep{gomez2006naps,wagner2004sleep,smith2003ingestion,djonlagic2009sleep,cai2009rem,Lewis2018How,durrant2015schema}, and protects against interference between memories~\citep{McDevitt2015REM}. In motor tasks, NREM is associated with the improvement of simple sequential tasks, while REM promotes the consolidation of complex motor tasks~\citep{King2017Sleepinga}. REM also selectively promotes consolidation of emotional memories~\citep{Baran2012Processing,Wagner2001Emotional}. This collection of evidence suggests NREM is associated with transfer and storage of recent experiences and REM is associated with organizing internal representations. This also coincides with NREM occurring more during early parts of night sleep, when transfer occurs, followed by REM sleep occurring predominately during the later part of night sleep, when integration and higher order memory manipulations occur. This is illustrated in Fig.~\ref{fig:biological-replay-types}.

%%%%% NREM VS REM FIGURE %%%%%

%%%%%%%%%%%%%%%%%%%%%%%%%%%%%%%%%%%%%%%%%%%%%%%%%%%%%%%%%%%%%%%%%%%%%%%%%%%%%%%%%%%%%%%%%%%%%%%%%%%%%%%%%%%%%%%%
\subsection{Replay Reflects Recent Memories}
\label{sec:org8a35045}

Since synaptic plasticity mechanisms that result in long-term changes are activity dependent~\citep{bi1998synaptic,markram1997regulation,abbott2000synaptic}, replay during sleep plays a critical role in long-term memory consolidation.

The first evidence of replay in the hippocampus was observed in the firing patterns of pairs of neurons~\citep{Pavlides1989Influences,Wilson1994Reactivation}. In these studies, the firing patterns of `place cells' in the hippocampus were measured during sleep. Since place cells are known to spike when the animal is in a particular location~\citep{OKeefe1978The}, it is possible to study neurons that are active during both sleep and recent waking experiences. A strong correlation was observed between the firing rates of place cell neurons during sleep to those observed during the waking task~\citep{Pavlides1989Influences,Wilson1994Reactivation}. Such replay of recent learning has been replicated across several studies~\citep{louie_temporally_2001,Davidson2009Hippocampal} and in other brain regions~\citep{peyrache_replay_2009,ji:2007:nat-neurosci:17173043}. Subsequent studies focus on the relative timing of different neurons, and identified a similarity in the temporal order of spiking between awake experiences and sleep~\citep{louie_temporally_2001,ji:2007:nat-neurosci:17173043}. 

Replay of recent memories in hippocampus decays over time, with the progressive reduction of correlations in the firing of neurons with recent experience over several sleep cycles~\citep{nadasdy1999replay}. There is also a reduction in the strength of replay during awake rest across days~\citep{karlsson_awake_2009}. The decline in replay of recent experiences was shown to occur relatively fast in hippocampus and prefrontal cortex, i.e., within a matter of hours~\citep{Kudrimoti1999Reactivation,Tatsuno2006Methodological}. The decline of replay in recent experiences during sleep is followed by the resetting of hippocampal excitability during REM sleep~\citep{grosmark:2012:neuron:22998869}, which may allow the hippocampus to encode new memories.

Replay of recent memories has also been observed in brain regions outside of the hippocampus. For example, the prefrontal cortex shows firing patterns similar to recent learning that are time-compressed~\citep{euston2007fast,peyrache_replay_2009}. Coordinated replay between hippocampus and visual cortex has also been observed~\citep{ji:2007:nat-neurosci:17173043}, which provides direct evidence for the transfer of memories from hippocampus to cortex. Recent procedural memories also result in the replay of task related activity during sleep. Scalp recordings during sleep resemble activity during recently learned motor tasks~\citep{Schonauer2014Strengthening}. Spiking activity from motor cortex during NREM sleep has firing patterns that are similar to recent learning~\citep{ramanathan2015sleep, gulati2017neural, Gulati2014Reactivation}. Further, replay during sleep was essential for solving the credit assignment problem by selectively increasing task related neuron activity and decreasing task unrelated neuron activity following sleep~\citep{gulati2017neural}. Taken together, these findings suggest recent memories are replayed by many different brain regions during sleep, some of which may relate to hippocampus, while others appear to originate locally or through other brain regions.

%%%%%%%%%%%%%%%%%%%%%%%%%%%%%%%%%%%%%%%%%%%%%%%%%%%%%%%%%%%%%%%%%%%%%%%%%%%%%%%%%%%%%%%%%%%%%%%%%%%%%%%%%%%%%%%%
\subsection{Selective Replay Enables Better Memory Integration}

Replay does not exactly replicate activity during waking experiences. An explicit demonstration of this effect was shown in a two choice maze task, where replay in rats corresponded to paths that were experienced and shortcuts that were never experienced~\citep{Gupta2010Hippocampal}. More recently, replay during sleep in the hippocampus and prefrontal cortex corresponded to separate activation of movement related and movement independent representations~\citep{Yu2017Distinct}. Likewise, commonly used methods for analyzing replay, e.g., template matching, principal component analysis, and independent component analysis, have shown high similarity, but not an exact match, between activation patterns during waking and sleep~\citep{lee_memory_2002,louie_temporally_2001, peyrache_replay_2009}.

Selective and partial replay are, in part, responsible for why replay is not an exact reconstruction of waking experience. Since sleep is time limited as compared to the time required to replay \emph{all} memories related to recent experiences, replaying only selected experiences can be more efficient for consolidation. Which experiences are selected for replay remains an open question. In \citep{schapiro2018human}, it was suggested that weakly learned information was replayed more frequently than other memories. Moreover, selective replay has also been shown to be motivated by fear~\citep{de2016awake} and reward~\citep{gruber2016post,murty2017selectivity,singer2009rewarded}. In \citep{mcclelland2020integration}, it was suggested that old experiences that overlap with new memories are in the most danger of being damaged by new learning and are preferentially replayed. Further, only certain sub-parts of the selected experiences are replayed. Such partial replay allows relevant memories with shared components to be blended, which could result in improved generalization. This was demonstrated in a recent experiment involving rats in various tasks, where coordinated partial replay of hippocampal and prefrontal cortical neurons represented generalization across different paths~\citep{Yu2018Specific}. Magnetoencephalography (MEG) studies in humans after training on tasks involving sequences with overlapping structure had activity evident of partial replays that resulted in generalization~\citep{Liu2019Human}. Partial replay has also been proposed to build cognitive schemata~\citep{lewis2011overlapping}. Thus, partial replay provides a mechanism for higher order memory operations, which is not a simple repetition of past experience.

Rewards received during tasks are strong modulators of replay during sleep. The temporal order of replay can be reversed when it is associated with a reward, and the strength of this reversal is correlated with the reward magnitude~\citep{Ambrose2016Reverse}. Reverse replay can be explained based on a form of spike time dependent synaptic plasticity (STDP), which allows for symmetric connections in both directions following sequential activation with rewards at the end of the sequence~\citep{pfeiffer2020content}. A series of human experiments further suggest that selective and partial replay in tasks involving reward allow humans to perform generalization and model-based reinforcement learning~\citep{Momennejad2020Learning}, which has inspired several algorithms in machine learning (see \citep{Caze2018Hippocampal} for a review). Another important function of selective and partial replay is in planning. Replay is shown to include activity sampled from past experiences as well as novel activity that corresponds to future possibilities~\citep{johnson2007neural} and random movements in place cells~\citep{stella2019hippocampal}. Experimental and theoretical studies have identified partial replay as a potential mechanism for exploring possible routes or facilitating goal-directed navigation~\citep{Foster2012Sequence,pfeiffer2013hippocampal}. For example, partial replay corresponded to different locations in the environment that could facilitate the reconstruction of unexplored paths and novel future trajectories~\citep{Olafsdottir2015Hippocampal,olafsdottir2018role}.

Since reverse replay begins with the state of the reward and spike sequences trace backwards, it has been proposed to be similar to backward planning from the goal in Markov decision processes~\citep{foster2017replay}. Thus, reverse replay could be an efficient way of estimating state values, which are critical in reinforcement learning.

%%%%%%%%%%%%%%%%%%%%%%%%%%%%%%%%%%%%%%%%%%%%%%%%%%%%%%%%%%%%%%%%%%%%%%%%%%%%%%%%%%%%%%%%%%%%%%%%%%%%%%%%%%%%%%%%
\subsection{Replay Generation and Coordination Across Brain Regions}

The exact mechanism for the spontaneous origin of replay during sleep and resting periods is not well understood. During sleep, there is a large change in the neuromodulatory tone for each sleep state across the entire brain~\citep{brown2012control,mccormick1992neurotransmitter,watson2010neuropharmacology}. Neuromodulatory levels determine a neuron's excitability and the strength of its synaptic connections. During NREM sleep, due to a reduction in acetylcholine and monoamine levels, there is an overall reduction in neuron excitability and an increase in excitatory connections. Further, increased extracellular GABA during NREM suggests an increase in inhibition during this state. The reduced excitability and heightened synaptic connections result in spontaneous activity during sleep~\citep{olcese2010sleep,krishnan2016cellular}. This reflects patterns of synaptic connectivity between neurons rather than the intrinsic state of neurons. Reactivation has also been observed in computational models of attractor networks~\citep{Shen1996Modeling}, where random activations initiate attractor dynamics that result in the replay of activity that formed the attractor. If synaptic changes reflect previous learning, such as a particular group of neurons co-activating or a sequential activation of neurons, then the activity generated during sleep follows or replays the activity from learning. This has been demonstrated in several studies involving recurrently connected thalamocortical networks~\citep{wei2016synaptic,Wei2018Differentialb,gonzalez2020can}. The studies demonstrated that replay helps to avoid interference between competing memories and results in a synaptic connection that reflects the combination of previous tasks~\citep{Golden2020Sleep,gonzalez2020can}. 

This mechanism is the basis of replay in computational models of attractor networks~\citep{crick1983function,robins1998catastrophic,robins1999consolidation}. Specifically, in \citep{crick1983function} attractors are randomly chosen for replay, facilitating the unlearning of them. Conversely, in \citep{robins1998catastrophic}, attractors are replayed for re-learning (e.g., via replay or generative replay). Both of these studies use Hopfield networks and are directly compared in \citep{robins1999consolidation}.

An important characteristic of sleep is the synchronization of firing across neurons that leads to oscillations (e.g., oscillations in the local field potential or Electroencephalography (EEG) signals). Replay during sleep was shown to co-occur with sleep oscillations. NREM sleep is characterized by several well-defined types of oscillations, found across a wide range of species from reptiles to humans, including sharp wave ripples (100-200 Hz) in the hippocampus~\citep{buzsaki1992high}, spindles  (7-14 Hz)~\citep{morison1941study} and slow ($<$ 1 Hz) oscillations~\citep{steriade2001natural,steriade1993novel} in the thalamocortical network~\citep{bazhenov2006thalamocortical}. Numerous studies have demonstrated that replay in the hippocampus is linked to the occurrence of sharp-wave ripples~\citep{nadasdy1999replay,foster2006reverse,Davidson2009Hippocampal,peyrache_replay_2009,buzsaki2015hippocampal}.  In the cortex, replay occurs during spindles and active states of slow oscillations~\citep{ramanathan2015sleep}. Indeed, oscillatory activities during NREM stage 2 sleep, including sleep spindles and slow waves, are strongly correlated with motor sequence memory consolidation~\citep{nishida2007daytime,Barakat2013Sleep}.  

There is evidence for the coordination between oscillations across the cortex and hippocampus~\citep{battaglia2004hippocampal,sirota2003communication,molle2006hippocampal,siapas1998coordinated}, suggesting that sleep rhythms can mediate the coordinated replay between brain regions. The nesting of ripples, spindles, and slow oscillations was reported in vivo~\citep{staresina2015hierarchical} and demonstrated in large-scale biophysical models~\citep{sanda2021bidirectional}. Coordinated replay is supported by simultaneous recordings from hippocampus and neocortex~\citep{ji:2007:nat-neurosci:17173043}. Taken together, these evidences strongly suggest that replay in the neocortex is modulated by the hippocampus during sleep. Such coordination is critical for the transfer of recent memories from hippocampus to cortex and also across cortical regions. This coordination leads to the formation of long-range connections and promotes associations across memories and modalities.

Properties of sleep oscillations influence replay and synaptic plasticity during sleep. The frequency of spiking during spindles is well suited for initiating spike timing dependent synaptic plasticity (STDP)~\citep{Sejnowski2000Why}. Both spindles and slow oscillations demonstrate characteristic spatio-temporal dynamics~\citep{muller2016rotating} and its properties determine synaptic changes during sleep~\citep{wei2016synaptic}.

%%%%%%%%%%%%%%%%%%%%%%%%%%%%%%%%%%%%%%%%%%%%%%%%%%%%%%%%%%%%%%%%%%%%%%%%%%%%%%%%%%%%%%%%%%%%%%%%%%%%%%%%%%%%%%%%
\subsection{Open Questions About Replay}
\label{sec:orgcd6414a}

There are several open questions about replay in biological networks that are far from being well understood. One of them is about the origin and functions of replay during REM sleep. REM sleep has been shown to play at least three intertwined roles~\citep{walker2010overnight}: 1) it \emph{unitizes} distinct memories for easier storage~\citep{haskins2008perirhinal,bader2010recognition,tibon2014associative,kuriyama2004sleep,ellenbogen2007human}, 2) it \emph{assimilates} new memories into existing networks~\citep{walker2002cognitive,stickgold1999sleep,dumay2007sleep}, and 3) it \emph{abstracts} high-level schemas and generalizations to unlearn biased or irrelevant representations~\citep{gomez2006naps,wagner2004sleep,smith2003ingestion,djonlagic2009sleep,cai2009rem,Lewis2018How,durrant2015schema}. REM sleep also facilitates creative problem-solving~\citep{Lewis2018How,baird2012inspired,cai2009rem}. 

While the majority of replay studies are from NREM sleep, some studies have shown replay during REM sleep~\citep{louie_temporally_2001,Eckert2020Neural,Kudrimoti1999Reactivation}. Early studies did not find a correlation in firing patterns during REM, which had been found in NREM sleep~\citep{Kudrimoti1999Reactivation}. However, \citep{louie_temporally_2001} identified replay during REM in hippocampal place cells similar to NREM. In the case of motor skill learning, reactivation was observed during both REM and NREM sleep. Moreover, replays during REM and NREM are interlinked, since replay during REM is correlated with replay during NREM from the previous night~\citep{Eckert2020Neural}. REM sleep was implicated in pruning newly-formed postsynaptic dendritic spines in the mouse motor cortex during development and motor learning and was also shown to promote the survival of new, learning-induced, spines that are important for the improvement of motor skills~\citep{li2017rem}. Together, these studies point to the important but still poorly understood role of REM sleep in memory and learning and suggest that the repetition of NREM and REM stages with different neuromodulatory states are critical for memory consolidation.

What is the meaning of replayed activity? While in some cases the replay faithfully replicates activity learned during awake, many evidences suggest that the content of replay is more than just a simple combination of past activities. As the brain learns new memories that may potentially try to allocate synaptic resources belonging to the old memories, sleep may not simply replay previously learned memories to avoid forgetting. Instead, sleep may change representations of the old memories by re-assigning different subsets of neurons and synapses to effectively orthogonalize memory representations and allow for overlapping populations of neurons to store multiple competing memories~\citep{gonzalez2020can}. In fact, orthogonalizing network representations was one of the earliest attempted solutions to catastrophic forgetting in artificial networks (see \citep{french1999catastrophic} for an overview).

One of the outstanding questions about replay involves the selection of replayed activity. As highlighted in previous works~\citep{mcclelland2020integration}, given the limited time period of sleep, only a subset of memories are selected for replay during sleep. This suggests that the neural activity during sleep is selected to maximize consolidation, while simultaneously preventing forgetting. Machine learning algorithms could optimize directly for which old memories to be replayed during consolidation to long-term memory; these ideas could inform neuroscience research. While there are major differences between the nature of activity in artificial and spiking networks, machine learning inspired replay methods could still provide insights about neural activity selection during sleep  replay.

%%%%%%%%%%%%%%%%%%%%%%%%%%%%%%%%%%%%%%%%%%%%%%%%%%%%%%%%%%%%%%%%%%%%%%%%%%%%%%%%%%%%%%%%%%%%%%%%%%%%%%%%%%%%%%%%
\section{Replay in Artificial Networks}
\label{sec:replay-artificial}

When a deep neural network can be trained in an offline setting with fixed training and testing datasets, gradient descent can be used to learn a set of neural weights that minimize a loss function. However, when the training set evolves in a non-stationary manner or the agent learns from a sequence of experiences, gradient descent updates to the network cause \emph{catastrophic forgetting} of previously learned knowledge~\citep{mccloskey1989,Abraham2005Memory}. This forgetting occurs because parametric models, including neural networks, assume that data is independent and identically distributed (iid). In offline settings, models can simulate the notion of iid experiences by shuffling data. However, in continual learning settings, the data stream is evolving in a non-iid manner over time, which causes catastrophic forgetting of previous knowledge.

Further, offline machine learning setups are unable to continually learn new data since they assume there are distinct periods of training versus evaluation, that the training and testing data come from the same underlying data distribution, and that all of the training data is available at once. When these assumptions are violated, the performance of neural networks degrades. The broad field of \emph{lifelong machine learning} seeks to overcome these challenges to continually train networks from evolving non-iid data streams. In addition to overcoming catastrophic forgetting, lifelong learning agents should be capable of using previous knowledge to learn similar information better and more quickly, which is known as forward knowledge transfer. This survey and much of the existing lifelong learning literature have focused on overcoming catastrophic forgetting using replay, however forward knowledge transfer is also an important aspect of lifelong learning that has received little attention~\citep{chaudhry2018riemannian,lopez2017gradient}, and should be studied in more detail.

Moreover, catastrophic forgetting occurs due to the stability-plasticity dilemma, which requires networks to keep weights of the network stable in order to preserve previous knowledge, but also keep weights plastic enough to learn new information. Three main types of methods for mitigating forgetting have been proposed~\citep{parisi2019continual,kemker2018forgetting,de2019continual}: 1) regularization schemes for constraining weight updates with gradient descent~\citep{kirkpatrick2017,aljundi2018memory,zenke2017continual,chaudhry2018riemannian,ritter2018online,serra2018overcoming,dhar2019learning,chaudhry2019efficient,lopez2017gradient}, 2) network expansion techniques for adding new parameters to a network to learn new information~\citep{rusu2016progressive,yoon2018lifelong,ostapenko2019learning,hou2018lifelong}, and 3) replay mechanisms for storing a representation of previous data to mix with new data when updating the network. Replay (or rehearsal) mechanisms have been shown to be the most effective of these approaches and are inspired by how the mammalian brain learns new information over time.

%%%%%%%%%%%%%%%%%%%%%%%%%%%%%%%%%%%%%%%%%%%%%%%%%%%%%%%%%%%%%%%%%%%%%%%%%%%%%%%%%%%%%%%%%%%%%%%%%%%%%%%%%%%%%%%%
\subsection{Replay in Supervised Learning}

The ability of agents to learn over time from non-stationary data distributions without catastrophic forgetting is known as continual learning. Within continual learning, there are two major paradigms in which agents are trained~\citep{parisi2019continual}. The first paradigm, known as incremental batch learning, is the most common~\citep{castro2018end,chaudhry2018riemannian,fernando2017,hou2019unified,kemker2018fearnet,kemker2018forgetting,rebuffi2016icarl,wu2019large,zenke2017continual}. In incremental batch learning, an agent is required to learn from a labelled dataset $\mathcal{D}$ that is broken into $T$ distinct \emph{batches}. That is, $\mathcal{D}=\bigcup_{t=1}^{T}B_{t}$, where each $B_{t}$ is a batch of data consisting of $N_{t}$ labelled training samples, i.e., $B_{t}=\{\left(\mathbf{x}_{i},y_{i}\right)\}_{i=1}^{N_{t}}$ with $\left(\mathbf{x}_{i},y_{i}\right)$ denoting a training sample. At time $t$, the agent is required to learn from batch $B_{t}$ by looping over the batch several times and making updates, before inference can be performed.

Although the incremental batch learning paradigm is popular in recent literature, it comes with caveats. Learning from batches is not biologically plausible and it is slow, which is not ideal for immediate learning. More specifically, mammals engage in resource constrained online learning from temporally correlated data streams, which is known as single pass online learning or \emph{streaming learning}. Streaming learning is a special case of incremental batch learning where the batch size is set to one ($N_{t}=1$) and the agent is only allowed a single epoch through the labelled training dataset~\citep{gama2010knowledge,gama2013evaluating}. This paradigm closely resembles how humans and animals immediately learn from real-time data streams and can use new knowledge immediately.

Replay is one of the earliest~\citep{hetherington1989there,ratcliff1990connectionist} and most effective mechanisms for overcoming forgetting in both the incremental batch~\citep{castro2018end,rebuffi2016icarl,wu2019large,hou2019unified,kemker2018fearnet,kemker2018forgetting} and streaming~\citep{hayes2019memory,hayes2019remind,chaudhry2019efficient,lopez2017gradient} paradigms. There are two ways in which replay has been used in artificial neural networks: partial replay and generative replay (pseudo-rehearsal). For partial replay, an agent will store either all or a subset of previously learned inputs in a replay buffer. It then mixes either all, or a subset of, these previous inputs with new samples and fine-tunes the network on this mixture. For example, several of the most successful models for incremental learning store a subset of previously learned raw inputs in a replay buffer~\citep{Gepperth2016,rebuffi2016icarl,lopez2017gradient,castro2018end,chaudhry2018riemannian,nguyen2018variational,hou2018lifelong,hayes2019memory,hou2019unified,wu2019large,lee2019overcoming,belouadah2019il2m,chaudhry2019efficient,riemer2018learning,aljundi2019online,aljundi2019gradient,Belouadah_2020_WACV,he2020incremental,zhao2020maintaining,kurle2019continual,chrysakis2020online,kim2020imbalanced,tao2020topology,douillard2020podnet}. However, replaying raw pixels is not biologically plausible. More recently, methods that store representations/features from the middle (latent) layers of a network for replay have been developed~\citep{hayes2019remind,iscen2020memory,caccia2019online,pellegrini2019latent}, which are more consistent with replay in the mammalian brain, as suggested by hippocampal indexing theory~\citep{Teyler2007The} (see Sec.~\ref{sec:replay-biology}). The challenge in using representational replay comes in choosing which hidden layer(s) to use replay features from. While choosing features from earlier layers in the network allows more of the network to be trained incrementally, early features usually have larger spatial dimensions and require more memory for storage. Choosing the ideal layer for representational replay remains an open question.

In contrast to storing previous examples explicitly, generative replay methods train a generative model such as an auto-encoder or a generative adversarial network (GAN) \citep{goodfellow2014generative} to generate samples from previously learned data~\citep{draelos2017neurogenesis,kemker2018fearnet,robins1995catastrophic,ostapenko2019learning,shin2017continual,he2018exemplar,french1997pseudo,atkinson1802pseudo,robins1996consolidation}. The first generative replay method was proposed in \citep{robins1995catastrophic}, where it was further suggested that these mechanisms might be related to memory consolidation during sleep in mammals. Similar to partial replay methods, generative replay methods can generate veridical inputs~\citep{shin2017continual,kemker2018fearnet,parisi2018lifelong,he2018exemplar,wu2018memory,ostapenko2019learning,abati2020conditional,liu2020mnemonics,Titsias2020Functional,Oswald2020Continual,ye2020learning} or mid-level CNN feature representations~\citep{van2020brain,lao2020continuous}. These generative approaches do not require the explicit storage of data samples, which could potentially reduce storage requirements and mitigate some concerns regarding privacy. However, the generator itself often contains as many parameters as the classification network, leading to large memory requirements. Additionally, generative models are notoriously difficult to train due to convergence issues and mode collapse, making these models less ideal for online learning. One advantage to using an unsupervised generative replay method is that the system could potentially be less susceptible to, but not completely unaffected by, catastrophic forgetting~\citep{gillies1991stability}. Additionally, generative replay is more biologically plausible as it is unrealistic to assume the human brain could store previous inputs explicitly, as is the case in partial replay. An overview of existing supervised replay methods for classification and their associated categorizations are in Fig.~\ref{fig:replay-methods}.

%% Machine learning methods figure %%%
\begin{figure}%[t]
    \centering
    \resizebox{\textwidth}{!}{
\begin{tikzpicture}
\matrix (Y) [matrix of nodes, nodes={draw, anchor=center, minimum height=1cm, minimum width=3.25cm, outer sep=0pt, fill=gray!30},
    column sep=-\pgflinewidth, row sep=-\pgflinewidth]{
    
    GeppNet & iCaRL & GEM & End-to-End & |[fill=green!30, text=black]| DGR & |[fill=green!30, text=black]| FearNet \\
    
    RWALK & VCL & AD & ExStream & |[fill=green!30, text=black]| GDM & |[fill=green!30, text=black]| ESGR \\
    
    LUCIR & BiC & GD & IL2M & |[fill=green!30, text=black]| MeRGAN & |[fill=green!30, text=black]| DGM \\
    
    A-GEM & MER & MIR & |[fill=blue!30, text=black]| REMIND & |[fill=green!30, text=black]| CCG & |[fill=green!30, text=black]| Mnemonics \\
    
    GSS & ScaIL & ILOS & |[fill=blue!30, text=black]| FA & |[fill=green!30, text=black]| FRCL & |[fill=green!30, text=black]| HNET \\
    
    WA & GRS & CBRS & |[fill=blue!30, text=black]| AQM & |[fill=green!30, text=black]| L-VAEGAN & |[fill=green!30, text=black]| \\
    
    PRS & TPCIL & PODNet & |[fill=blue!30, text=black]| AR1* & |[fill=red!30, text=black]|  BI-R & |[fill=red!30, text=black]| DAFR\\
    };
\draw (-3, 4) node {\Large Raw Replay Methods};
\draw (6.5, 4) node {\Large Generative Replay Methods};
\end{tikzpicture}
}
    \caption{Categorizations of supervised artificial replay algorithms: \textcolor{Gray}{Veridical Replay}; \textcolor{Blue}{Representational Replay}; \textcolor{Green}{Generative Veridical Replay}; \textcolor{Red}{Generative Representational Replay}. See Appendix (Table~\ref{table:method-cites}) for algorithm citations.}
    \label{fig:replay-methods}
\end{figure}

In addition to the aforementioned models that perform replay by storing a subset of previous inputs, there have been several models that use replay in conjunction with other mechanisms to mitigate forgetting, such as regularizing parameter updates. For example, the Gradient Episodic Memory (GEM)~\citep{lopez2017gradient} and Averaged-GEM~\citep{chaudhry2019efficient} models store a subset of previous inputs to use with a gradient regularization loss. Similarly, the Meta-Experience Replay model~\citep{riemer2018learning} and Variational Continual Learning model~\citep{nguyen2018variational} use experience replay in conjunction with meta-learning and Bayesian regularization techniques, respectively.

For both partial replay and generative replay approaches, the agent must decide \emph{what} to replay. In \citep{chaudhry2018riemannian}, four selection strategies are compared for storing a small set of previous exemplars to use with a regularization approach. Namely, they compare uniform random sampling, storing examples closest to class decision boundaries, storing examples with the highest entropy, and storing a mean vector for each class in deep feature space. While they found storing a representative mean vector for each class performed the best, uniform random sampling performed nearly as well with less compute. In \citep{aljundi2019online}, samples that would be the most interfered with after network updates are replayed to the network, i.e., those samples for which performance would be harmed the most by parameter updates. In their experiments, the authors found that replaying these interfered samples improved performance over randomly replaying samples. Similarly, in \citep{aljundi2019gradient} sample selection is formulated as a constrained optimization problem, which maximizes the chosen sample diversity. The authors of \citep{aljundi2019gradient} further propose a greedy sampling policy as an alternative to the optimization and find that both sample selection policies improve performance over random selection. Similarity scores have also been used to select replay samples~\citep{mcclelland2020integration}. While selective replay has demonstrated promising results in some small-scale settings, several large-scale studies have found that uniform random sampling works surprisingly well~\citep{hayes2019remind,wu2019large}, achieving almost the same performance as more complicated techniques, while requiring less compute. The sample selection problem is also closely related to active learning strategies~\citep{cohn1994improving,lin2017active,wang2016cost,settles2009active,yoo2019learning,wei2015submodularity}, with the most common selection methods using uncertainty sampling~\citep{Lewis94asequential,culotta2005reducing,scheffer2001active,dagan1995committee,dasgupta2008hierarchical}. 

In addition to improving accuracy, selective replay can also facilitate better sample efficiency, i.e., the network requires fewer samples to learn new information. Sample efficiency has been studied for continual learning~\citep{davidson2020sequential}. In \citep{davidson2020sequential}, the authors found that a convolutional neural network required fewer training epochs to reach a target accuracy on a new task after having learned other visually similar tasks. These findings are closely related to the multi-task learning literature, where the relationship between task similarity and network performance in terms of accuracy and time has been studied~\citep{zhang2017survey,ruder2017overview,standley2020tasks,liu2019end,kendall2018multi}.

While the majority of supervised learning literature has focused on overcoming forgetting in feed-forward or convolutional neural networks, there has also been work focused on using replay to mitigate forgetting in recurrent neural networks~\citep{parisi2018lifelong,sodhani2020toward}. In \citep{parisi2018lifelong}, a self-organizing recurrent network architecture consisting of a semantic memory and an episodic memory is introduced to replay previous neural reactivations. In \citep{sodhani2020toward}, a network expansion technique is combined with gradient regularization and replay in a recurrent neural network to mitigate forgetting. 

Further, in \citep{hayes2019remind,greco-2019-psycholinguistics}, replay is used as an effective mechanism to mitigate forgetting for the problem of visual question answering, where an agent must answer natural language questions about images. Similarly, replay has been used for continual language learning~\citep{de2019episodic}. Replay has also been used to perform continual semantic segmentation of medical images~\citep{ozdemir2018learn,ozdemir2019extending}, remote sensing data~\citep{tasar2019incremental,wu2019ace}, and on standard computer vision benchmarks~\citep{cermelli2020modeling}. In \citep{acharya2020rodeo,liu2020continual}, replay is used to mitigate forgetting for a continual object detection approach. Replay approaches have also been explored in continual learning for robotics~\citep{lesort2020continual,feng2019challenges}.

%%%%%%%%%%%%%%%%%%%%%%%%%%%%%%%%%%%%%%%%%%%%%%%%%%%%%%%%%%%%%%%%%%%%%%%%%%%%%%%%%%%%%%%%%%%%%%%%%%%%%%%%%%%%%%%%
\subsection{Replay in Reinforcement Learning}

Experience replay has also been widely used in reinforcement learning~\citep{mnih2015human,mnih2013playing,van2016deep,lillicrap2015continuous,lin1992self,adam2011experience,foerster2017stabilising,kapturowski2018recurrent,atkinson2021pseudo}. As in supervised classification, experience replay in reinforcement learning is inspired by the interplay between memory systems in the mammalian brain and its biological plausibility has been discussed in \citep{Schaul2016PrioritizedER,hassabis2017neuroscience}. The overall goal of reinforcement learning is to train an agent to appropriately take actions in an environment to maximize its reward, which is a naturally realistic setup as compared to existing supervised classification setups. In online reinforcement learning, an agent is required to learn from a temporally correlated stream of experiences. However, the temporal correlation of the input stream is not independent and identically distributed and violates the assumptions of conventional, gradient-based optimization algorithms typically used for updating agents, resulting in catastrophic forgetting. \citet{lin1992self} proposed experience replay as a method for creating independent and identically distributed batches of data for an agent to learn from, while additionally allowing the agent to store and replay experiences that are rarely encountered. Specifically, the Deep Q-Network (DQN)~\citep{mnih2013playing,mnih2015human} performed experience replay using a sliding window approach where a uniformly selected set of previous transitions was replayed to the agent. While the random experience selection policy helped stabilize training of the DQN, prioritized experience replay~\citep{Schaul2016PrioritizedER} has been shown to be more effective and efficient. Prioritized experience replay is based on the assumption that some transitions between experiences may be more surprising to the agent and additionally that some experiences might not be immediately relevant to an agent and should be replayed at a later point during training~\citep{schmidhuber1991curious}. 

In \citep{Schaul2016PrioritizedER}, prioritized experience replay was performed based on the magnitude of an experience's temporal-difference (TD) error, which measures an agent's learning progress and is consistent with biological findings~\citep{singer2009rewarded,mcnamara2014dopaminergic}. However, using TD error alone can result in less diverse samples being replayed and must be combined with an importance-based sampling procedure. In \citep{isele2018selective}, the authors compared four experience selection strategies to augment a first-in first-out queue, namely: TD error, absolute reward, distribution matching based on reservoir sampling, and state-space coverage maximization based on the nearest neighbors to an experience. They found that experience selection based on TD error and absolute reward did not work well in mitigating forgetting, while selection based on distribution matching and state-space coverage had comparable performance to an unlimited replay buffer. Although replay selection strategies have not shown as much benefit for the supervised learning scenario, they have significantly improved the performance and efficiency of training in reinforcement learning agents~\citep{moore1993prioritized}.

In standard prioritized experience replay, each experience is typically associated with a single goal (reward). In contrast, hindsight experience replay~\citep{andrychowicz2017hindsight} allows experiences to be replayed with various rewards, which has several advantages. First, it allows learning when reward signals are sparse or binary, which is a common challenge in reinforcement learning agents. Overcoming sparse reward signals leads to sample efficiency. More interestingly, hindsight experience replay can serve as a form of curriculum learning~\citep{bengio_curriculum_2009} by structuring the rewards such that they start off simple and grow increasingly more complex during training. Curriculum learning has been shown to speed up the training of neural networks, while also leading to better generalization~\citep{bengio_curriculum_2009,graves_automated_2017,hunziker2018teaching,zhou2018minimax,fan2018learning,achille2018critical}. Additionally, curriculum learning is important for cognitive development in humans~\citep{lenneberg1967biological,senghas2004children}.

In addition to experience replay alone, several methods have also incorporated other brain-inspired mechanisms into their online reinforcement learning agents. For example, \citep{pritzel2017neural,lengyel2008hippocampal,blundell2016model} take inspiration from the role the hippocampus plays in making decisions to develop agents that learn much faster than other approaches. In \citep{chen2019toward}, the authors propose using only the raw environment pixel inputs for their agent in a trial-and-error scenario, which closely resembles how humans learn about and navigate their environments. In \citep{lake2017building}, it is argued that human brains are similar to model-free reinforcement learning agents for discrimination and associative learning tasks. 

%%%%%%%%%%%%%%%%%%%%%%%%%%%%%%%%%%%%%%%%%%%%%%%%%%%%%%%%%%%%%%%%%%%%%%%%%%%%%%%%%%%%%%%%%%%%%%%%%%%%%%%%%%%%%%%%
\subsection{Replay in Unsupervised Learning}

Although replay has been more extensively explored in supervised classification and reinforcement learning, it has also been explored in unsupervised learning settings~\citep{lesort2019generative,wu2018memory}. For example, replay has been explored in continual learning of GANs for image and scene generation. Specifically, the Exemplar-Supported Generative Reproduction model~\citep{he2018exemplar} uses a GAN to generate pseudo-examples for replay during continual learning, while the Dynamic Generative Memory model~\citep{ostapenko2019learning}, the Deep Generative Replay model~\citep{shin2017continual}, the Memory Replay GAN model~\citep{wu2018memory}, and the Closed-Loop GAN model~\citep{rios2018closed} are all used to continually learn to generate images and scenes. Continual learning with replay in GANs has also been used for reinforcement learning~\citep{caselles2019s}. Moreover, unsupervised learning techniques such as auto-encoders and GANs are widely used to generate replay samples in supervised learning algorithms~\citep{draelos2017neurogenesis,kemker2018fearnet}.

%%%%%%%%%%%%%%%%%%%%%%%%%%%%%%%%%%%%%%%%%%%%%%%%%%%%%%%%%%%%%%%%%%%%%%%%%%%%%%%%%%%%%%%%%%%%%%%%%%%%%%%%%%%%%%%%
\section{Juxtaposing Biological and Artificial Replay}
\label{sec:artificial-vs-biological-replay}

Recently, machine learning researchers have tried to bridge some of the differences between biological replay and artificial replay. For example, several methods using representational replay~\citep{hayes2019remind,caccia2019online,pellegrini2019latent,iscen2020memory} or generative representational replay~\citep{van2020brain,lao2020continuous}, instead of veridical (raw pixel) replay, have been proposed to improve continual learning performance. Moreover, \citep{mcclelland2020integration} uses similarity scores in selecting which samples to replay based on evidence of replay in the hippocampus and cortex. Further, \citep{tadros2020biologically,tadros2019biologically,krishnan2019biologically} implement a sleep-inspired mechanism in a converted spiking neural network to reduce catastrophic forgetting. More recently, \citep{van2020brain} incorporated several brain-inspired mechanisms into their artificial network including feedback connections, context gating mechanisms, and generative representational replay.

However, many replay algorithms still differ from how humans learn and assimilate new information. For example, few existing techniques use Hebbian or error-based learning~\citep{tao2020topology,parisi2018lifelong} and most rely on supervised labels during training. Moreover, epigenetic tagging mechanisms in medial prefrontal cortex have largely been ignored by existing artificial network approaches and some approaches largely focus on replay during waking hours instead of replay during sleep~\citep{hayes2019memory,hayes2019remind}. In biological networks, replay happens both independently and concurrently in several different brain regions, whereas artificial replay implementations only perform replay at a single layer within the neural network. Furthermore, many existing artificial replay implementations 1) do not purge their memory buffer, which is not consistent with biology~\citep{nadasdy1999replay,karlsson_awake_2009} and 2) do not have a notion of waking (streaming/online) learning.

While selective experience replay has yielded significant performance gains in reinforcement learning, uniform random sampling still works well and is widely used in supervised classification, which is not consistent with how memories are selectively replayed in the brain. It is biologically infeasible to store everything a mammal encounters in its lifetime and likewise, it is not ideal for machine learning agents to store all previous data. In the case of partial replay, several different strategies have been explored for prioritizing what memories should be replayed~\citep{chaudhry2018riemannian,aljundi2019gradient,aljundi2019online,mcclelland2020integration}. While there have been several replay selection methods proposed, many existing works have found uniform random sampling of previous memories to work well, especially for large-scale problems~\citep{chaudhry2018riemannian,wu2019large,hayes2019remind}. While sampling strategies have not demonstrated significant success for supervised learning problems, the reinforcement learning community has seen more benefit from these approaches, e.g., prioritized experience replay~\citep{Schaul2016PrioritizedER} and hindsight replay~\citep{andrychowicz2017hindsight}. Exploring selective replay strategies in machine learning could help inform biologists about what might be replayed in the brain. Further, the efficiency of selective replay in machine learning has largely been ignored with most researchers developing selective replay methods that only yield better performance. By studying selective replay techniques that are efficient, the agent could potentially learn information better and more quickly, which is closely related to forward knowledge transfer in humans. Moreover, humans generate novel memories that are not generated from external world inputs during REM sleep~\citep{Lewis2018How}. Exploring schemes to generate novel memories in machine learning could further improve performance.

In machine learning and computer vision, there have been several models inspired by CLS theory in biological networks~\citep{Gepperth2016,french1997pseudo,ans1997avoiding,kemker2018fearnet,robins1995catastrophic,robins1996consolidation}. All of these models have, or presuppose, a fast-learning hippocampal-inspired network and consolidate knowledge to a medial prefrontal cortex network that learns more slowly. Further, \citep{Gepperth2016,kemker2018fearnet,draelos2017neurogenesis} integrate neurogenesis into their models where new neurons are created to form new memories. Such neurogenesis-inspired mechanisms have been studied experimentally in biology~\citep{kumar2020sparse,deng2010new,aimone2014regulation,aimone2011resolving}. Several of these CLS-inspired models focus on using generative replay to generate new inputs during training~\citep{french1997pseudo,ans1997avoiding,kemker2018fearnet,robins1995catastrophic}, instead of storing raw inputs explicitly~\citep{Gepperth2016}. However, the vast majority of existing replay approaches in artificial neural networks replay raw pixel inputs~\citep{hou2019unified,rebuffi2016icarl,castro2018end,wu2019large}. There have been a few approaches that store high-level feature representations (feature maps) of inputs instead of using generative replay~\citep{hayes2019memory,hayes2019remind,pellegrini2019latent,caccia2019online}, which is more biologically plausible than replay from raw pixels. While there have been several models inspired by CLS theory~\citep{Gepperth2016,french1997pseudo,ans1997avoiding,kemker2018fearnet,robins1995catastrophic}, many existing replay approaches have only focused on modeling medial prefrontal cortex directly and do not have a fast learning network. Moreover, \citep{kemker2018fearnet} is the only CLS-inspired model that integrates a non-oracle basolateral amygdala network for decision-making during inference. Lastly, none of the aforementioned CLS-inspired models use information from the neocortex-inspired network to influence training of the hippocampal-inspired network, whereas the neocortex influences learning in the hippocampus and vice-versa in biological networks.

Further, CLS theory assumes that different awake and sleep states correspond to periods of encoding memories in hippocampus and the subsequent transfer of memories from hippocampus to cortex. This suggests that artificial neural networks could benefit from the inclusion of explicit awake and sleep states, inspired by the mammalian brain. Moreover, one open question in biology involves what happens to memories after they have been consolidated from the hippocampus to the neocortex. These memories in hippocampus could be erased entirely, or they could still be encoded in the hippocampus, but never reactivated again. While this is an open question in biology, its exploration in machine learning could inform the neuroscientific community and lead to new discoveries.

While the CLS memory model (i.e., fast learning in hippocampus followed by slow learning in the cortex) is widely accepted as a core principle of how the brain learns declarative memories, it is likely not the only memory model the brain uses. Indeed, procedural, presumably hippocampus-independent memories, e.g., some motor tasks~\citep{fogel2006learning} can be learned without forgetting old skills and replayed during REM sleep when the hippocampus is effectively disconnected from the neocortex. Even if the hippocampus may be important for early phases of motor learning, subsequent training and replay rely on motor cortex and striatal networks~\citep{lemke2019emergent}. Therefore, while typical machine learning replay approaches interleave new training data with old knowledge from hippocampus-like networks, the biological cortex is capable of replaying old traces on its own. For example, very few researchers have explored ``self-generated'' replay as a mechanism to protect old knowledge for continual learning~\citep{tadros2020biologically,tadros2019biologically,krishnan2019biologically}. Another alternative to CLS theory is the idea of using both fast and slow weights between units for each connection in the network~\citep{hinton1987using}. While the fast weights are used to learn new information, the slow weights can be used for generative replay~\citep{robins1997maintaining,robins2004sequential}.

Beyond CLS theory, more interesting computational models of HC have been explored. For example, \citep{kali2004off} proposed a model where the HC and cortex were modelled as a lookup-table and a restricted-Boltzmann-machine, respectively. In this model, the HC network played a critical role in memory retrieval, beyond serving as an offline replay buffer, which is how HC is commonly modelled in modern neural network implementations. The authors further discuss index maintenance and extension. More recently, \citep{whittington2020tolman} proposed the Tolman Eichenbaum Machine, where the HC network performs space and relational inference. Replay in the Tolman Eichenbaum Machine allowed for the organization of sequences into structures that could facilitate abstraction and generalization.

Another critical difference between biological and artificial implementations of replay is the notion of regularization. In biological networks, normalization and synaptic changes co-occur with replay~\citep{chauvette2012sleep,tononi2014sleep}. However, in artificial networks, regularization and replay approaches for mitigating catastrophic forgetting have largely been explored independently. While some deep learning methods combine replay and regularization~\citep{chaudhry2019efficient,lopez2017gradient}, each mechanism operates largely without informed knowledge of the other, unlike the co-occurrence and direct communication between the two mechanisms in biology. By integrating the two mechanisms with more communication in artificial networks, performance could be improved further and each mechanism could potentially strengthen the other component. For example, replay informed regularization could help strengthen connections specific to a particular memory, while regularization informed replay could help identify which samples to replay that will enable more transfer or less forgetting.

%%%%%%%%%%%%%%%%%%%%%%%%%%%%%%%%%%%%%%%%%%%%%%%%%%%%%%%%%%%%%%%%%%%%%%%%%%%%%%%%%%%%%%%%%%%%%%%%%%%%%%%%%%%%%%%%
\section{Conclusions}
\label{sec:conclusions}

Although humans and animals are able to continuously acquire new information over their lifetimes without catastrophically forgetting prior knowledge, artificial neural networks lack these capabilities~\citep{parisi2019continual,kemker2018forgetting,de2019continual}. Replay of previous experiences or memories in humans has been identified as the primary mechanism for overcoming forgetting and enabling continual knowledge acquisition~\citep{walker2004sleep}. While replay-inspired mechanisms have enabled artificial networks to learn from non-stationary data distributions, these mechanisms differ from biological replay in several ways. Moreover, current artificial replay implementations are computationally expensive to deploy. In this paper, we have given an overview of the current state of research in both artificial and biological implementations of replay and further identified several gaps between the two fields. By incorporating more biological mechanisms into artificial replay implementations, we hope deep networks will exhibit better transfer, abstraction, and generalization. Further, we hope that advancing replay in artificial networks can inform future neuroscientifc studies of replay in biology.

%%%%%%%%%%%%%%%%%%%%%%%%%%%%%%%%%%%%%%%%%%%%%%%%%%%%%%%%%%%%%%%%%%%%%%%%%%%%%%%%%%%%%%%%%%%%%%%%%%%%%%%%%%%%%%%%
\subsection*{Acknowledgments}
TH and CK were supported in part by the DARPA/SRI Lifelong Learning Machines program [HR0011-18-C-0051], AFOSR grant [FA9550-18-1-0121], and NSF award \#1909696. MB and GK were supported in part by the Lifelong Learning Machines program from DARPA/MTO [HR0011-18-2-0021], ONR grant [N000141310672], NSF award [IIS-1724405], and NIH grant [R01MH125557]. The views and conclusions contained herein are those of the authors and should not be interpreted as representing the official policies or endorsements of any sponsor. One author in this study, CK, was employed at a commercial company, Paige, New York during the preparation of this manuscript. This company played no role in the sponsorship, design, data collection and analysis, decision to publish, or preparation of the manuscript. We thank Anthony Robins and our anonymous reviewers and editors for their helpful comments and suggestions to improve this manuscript.

%%%%%%%%%%%%%%%%%%%%%%%%%%%%%%%%%%%%%%%%%%%%%%%%%%%%%%%%%%%%%%%%%%%%%%%%%%%%%%%%%%%%%%%%%%%%%%%%%%%%%%%%%%%%%%%%
\section*{Appendix}
\label{appendix}

\begin{center}
\small 
\rowcolors{0}{white}{light-gray}
\begin{longtable}{|c|c|}
\caption{Replay algorithm citations from Fig.~\ref{fig:replay-methods}.}
\label{table:method-cites}\\

\hline \rowcolor{white}
\textbf{\textsc{Algorithm}} & \textbf{\textsc{Citation}} \\
\hline
\endfirsthead

\multicolumn{2}{c}%
{\tablename\ \thetable\ -- \textit{Continued from previous page}} \\
\hline \rowcolor{white}
\textbf{\textsc{Algorithm}} & \textbf{\textsc{Citation}} \\
\hline
\endhead
\hline \multicolumn{2}{r}{\textit{Continued on next page}} \\
\endfoot
\hline
\endlastfoot
\hline
\multicolumn{2}{|l|}{\textbf{\textit{Veridical Replay}}}\\
\hline
GeppNet & \citep{Gepperth2016} \\
iCaRL & \citep{rebuffi2016icarl} \\
GEM & \citep{lopez2017gradient} \\
End-to-End & \citep{castro2018end} \\
RWALK & \citep{chaudhry2018riemannian} \\
VCL & \citep{nguyen2018variational} \\
AD & \citep{hou2018lifelong} \\
ExStream & \citep{hayes2019memory} \\
LUCIR & \citep{hou2019unified} \\
BiC & \citep{wu2019large} \\
GD & \citep{lee2019overcoming} \\
IL2M & \citep{belouadah2019il2m} \\
A-GEM & \citep{chaudhry2019efficient} \\
MER & \citep{riemer2018learning} \\
MIR & \citep{aljundi2019online} \\
GSS & \citep{aljundi2019gradient} \\
ScaIL & \citep{Belouadah_2020_WACV} \\
ILOS & \citep{he2020incremental} \\
WA & \citep{zhao2020maintaining} \\
GRS & \citep{kurle2019continual} \\
CBRS & \citep{chrysakis2020online} \\
PRS & \citep{kim2020imbalanced} \\
TPCIL & \citep{tao2020topology} \\
PODNet & \citep{douillard2020podnet} \\
\hline
\multicolumn{2}{|l|}{\textbf{\textit{Representational Replay}}}\\
\hline
REMIND & \citep{hayes2019remind} \\
FA & \citep{iscen2020memory} \\
AQM & \citep{caccia2019online} \\
AR1* & \citep{pellegrini2019latent} \\
\hline
\multicolumn{2}{|l|}{\textbf{\textit{Generative Veridical Replay}}}\\
\hline
DGR & \citep{shin2017continual} \\
FearNet & \citep{kemker2018fearnet} \\
GDM & \citep{parisi2018lifelong} \\
ESGR & \citep{he2018exemplar} \\
MeRGAN & \citep{wu2018memory} \\
DGM & \citep{ostapenko2019learning} \\
CCG & \citep{abati2020conditional} \\
Mnemonics & \citep{liu2020mnemonics} \\
FRCL & \citep{Titsias2020Functional} \\
HNET & \citep{Oswald2020Continual} \\
L-VAEGAN & \citep{ye2020learning} \\
\hline
\multicolumn{2}{|l|}{\textbf{\textit{Generative Representational Replay}}}\\
\hline
BI-R & \citep{van2020brain} \\
DAFR & \citep{lao2020continuous} \\
\end{longtable}
\end{center}

\bibliographystyle{apa}
\bibliography{references.bib}

\end{document}